\documentclass[aps,showpacs,showkeys,twocolumn]{revtex4}
\usepackage{mathrsfs}
\usepackage{amssymb}
\usepackage{amsmath}
\usepackage{bm}
\usepackage{graphics}
\usepackage{natbib}

\begin{document}
\title{Spin-dependent electron transport through a parallel double-quantum-dot structure}
\author{Weijiang Gong$^1$}\email[Corresponding author.
Fax: +086-024-8367-6883; phone: +086-024-8367-8327; Email address: ]
{weijianggong@gmail.com}
\author{Yisong Zheng$^2$} \affiliation{1. College of sciences, Northeastern University, Shenyang
110004, China \\2. Department of physics, Jilin University,
Changchun 130023, China}
\date{\today}

\begin{abstract}
Electron transport properties in a parallel double-quantum-dot
structure with three-terminals are theoretically studied. By
introducing a local Rashba spin-orbit coupling, we find that an
incident electron from one terminal can select a specific terminal
to depart from the quantum dots according to its spin state. As a
result, spin polarization and spin separation can be simultaneously
realized in this structure. And spin polarizations in different
terminals can be inverted by tuning the structure parameters. The
underlying quantum interference that gives rise to such a result is
analyzed in the language of Feynman paths for the electron
transmission.
\end{abstract}
 \maketitle

\bigskip

\section{Introduction}
Spintronics is one of the most attractive investigation frontier
both for the theoretical and experimental aspects due to the
potential application of nano-devices, in which generating and
controlling a spin-polarized current through a mesoscopic system is
the major issue as demonstrated in a variety of relevant
works\cite{Das,science}. One of the feasible techniques to achieve
spin-polarized current is electrical spin injection, but the
efficiency through ferromagnetic/ nonmagnetic semiconductor
junctions is usually small due to the conductivity
mismatch\cite{Das2,Das3,Filip}. An alternative method is the optical
spin orientation technique\cite{Hbner}, which is difficult to
integrate with electronic devices.

\par
Spin-orbit(SO) coupling is an important mechanism that influences
the electron spin state, since it couples the spin degree of freedom
of an electron to its orbital motion and vice versa. In particular,
in low-dimensional structures Rashba SO interaction comes into play
by exerting an electric potential to destroy the symmetry of space
inversion in an arbitrary spatial
direction\cite{Rashba,Rashba2,Rashba3,Sun}. Thus, based on the
properties of Rashba interaction, electric control and manipulation
of the spin state is feasible. Since Datta and Das proposed the
concept of spin-field-effect transistor\cite{Datta}, effects of SO
interaction on the electronic transport properties have been paid
much attention\cite{Kato,Sinova,Xie,Serra,Chi}. During the past two
decades, a great number of studies have been devoted to improve the
efficiency of spin polarization in the transport system based on the
SO interaction but not under magnetic field or coupled FM
leads\cite{Das}. In the spin Hall devices, it has been observed that
the pure spin current is induced in the transverse direction by the
SO interaction by applying a longitudinal electric field in the
two-dimensional electron (hole) system\cite{Kato,Sinova}. Very
recently, Rashba interaction has been introduced to quantum dot (QD)
systems, e.g., the Rashba interaction is locally applied to one QD
in one arm of an Aharonov-Bohm (AB)
interferometer\cite{Sun,Xie,Chi}. In these structures, with the
interplay of the magnetic field and Rahsba SO interaction,
remarkable spin polarization comes into being during the electron
transport process.
\par
In the present work, we introduce Rashba SO interaction to act
locally on one QD of a double QD AB ring, in which an additional
lead is laterally coupled to another QD. In comparison with a single
QD structure, such double QDs have more tunable parameters, and
provide more Feynman paths for electron transmission. Furthermore,
Rashba interaction can bring about the spin-dependent phase.
Accordingly, it is expected that quantum interference in such a
double QD structure can give rise to the spin-related electronic
transport properties. The most interesting result we obtain is that
an incident electron from one terminal can select a specific
terminal to depart from the QDs according to its spin state. In
other words, such a structure can realize the functions of spin
polarization and spin separation simultaneously. And these functions
can be tuned by the structure parameters. We find that the Rashba
interaction and the terminal triplet play the crucial roles in
creating such a feature.

\par
The rest of the paper is organized as follows. In Sec. II, the
electron Hamiltonian of second-quantization including the Rashba SO
interaction in the double-QD structure is introduced first. Then a
formula for linear conductance is derived by means of the
nonequilibrium Green function technique. In Sec. III, the calculated
results about the conductance spectra are shown. Then a discussion
concerning the spin polarization and separation is given. Finally,
the main results are summarized in Sec. IV.

\section{Model}\label{theory}
 The parallel double QD structure that we consider is illustrated in Fig. \ref{structure}. Apart from the
left and right leads, another lead is applied to couple laterally to
QD-2. A gate voltage is applied on QD-1 to induce a local Rashba SO
interaction. The Hamiltonian to describe the electronic motion in
the structure reads
\begin{equation}
H=H_{C}+H_{D}+H_{T}.   \label{1}
\end{equation}
The first term is the Hamiltonian for the noninteracting electrons
in three leads:
\begin{equation}
H_{C}=\sum_{\sigma, k,\alpha}\varepsilon
_{k\alpha}c_{k\alpha\sigma}^\dag c_{k\alpha\sigma },\label{2}
\end{equation}
where $c_{k\alpha\sigma}^\dag$ $( c_{k\alpha\sigma})$ is an operator
to create (annihilate) an electron of the continuous state
$|k\alpha,\sigma\rangle$ in the lead-$\alpha$($\alpha=L,R,D$), and
$\varepsilon _{k\alpha}$ is the corresponding single-particle
energy. $\sigma=\uparrow,\downarrow$ denotes the spin index. The
second term describes electron in double QDs, which takes a form as
\begin{equation}
H_{D}=\sum_{j,\sigma}\varepsilon _{j}d_{j\sigma}^\dag
d_{j\sigma}+\sum_{j}U_jn_{j\uparrow}n_{j\downarrow},\label{3}
\end{equation}
where $d^{\dag}_{j\sigma}$ $(d_{j\sigma})$ is the creation
(annihilation) operator of electron in QD-j($j=1,2$),
$\varepsilon_{j}$ denotes the electron level in the corresponding
QD. $U_j$ represents the intradot Coulomb repulsion, and the
interdot electron interaction is ignored since it is usually much
smaller than the intradot one due to the screening effect. The last
term in the total Hamiltonian describes the electron tunneling
between the leads and QDs, which is given by
\begin{eqnarray}
H_{T} &&=\sum_{\sigma,j,k}(V_{jL\sigma}d_{j\sigma}^\dag
c_{kL\sigma}+V_{jR\sigma}d_{j\sigma}^\dag
c_{kR\sigma})\nonumber\\
&&+\sum_{\sigma,k}V_{2D\sigma}d_{2\sigma}^\dag c_{kD\sigma}+\mathrm
{H.c.}, \label{4}
\end{eqnarray}
where $V_{j\alpha\sigma}$ denotes the QD-lead coupling coefficients.
They take the forms as: $V_{1L\sigma}=Ve^{i\phi/4+i\sigma\varphi}$,
$V_{1R\sigma}^*=Ve^{i\phi/4+i\sigma\varphi}$,
$V^*_{2L\sigma}=V_{2R\sigma}=Ve^{i\phi/4}$, and $V_{2D\sigma}=V_D$.
This implies that the couplings between the L and R leads and the
QDs have the uniform strength $V$ but different phases. The coupling
between lead-D and QD-2 can have the different strength. The phase
$\phi$ is associated with the magnetic flux $\Phi$ threading the
system by a relation $\phi=2\pi\Phi/\Phi_{0}$, where $\Phi_{0}=h/e$
is the flux quantum. Besides, the spin-dependent phase $\sigma\phi$
arises from the electron spin precession induced by the Rashba SO
coupling\cite{Sun}.
\par
Next we turn to discuss the electron transport through this
structure. With the Green function technique, the current flow in
lead-$\beta$ can be written as\cite{Meir1,Meir2}
\begin{equation}
J_{\beta}=\sum_\sigma
J_{\beta\sigma}=\frac{e}{h}\sum_{\alpha\sigma}\int d\omega
T_{\beta\alpha,\sigma}(\omega)[f_\beta(\omega)-f_{\alpha}(\omega)],\label{current}
\end{equation}
where
$T_{\beta\alpha,\sigma}(\omega)=\mathrm{Tr}[\bm{\Gamma}^\beta_{\sigma}
 \bm{G}^r_{\sigma}(\omega)\bm{\Gamma}^\alpha_{\sigma}\bm{G}^a_{\sigma}(\omega)]$
is the transmission function, describing electron tunneling ability
between lead-$\beta$ and lead-$\alpha$ via double QDs, and
$f_\alpha(\omega)$ is the Fermi distribution function in
lead-$\alpha$. The QD-lead coupling matrix
$[\bm{\Gamma}^\alpha_\sigma]$ is defined as
$[\bm{\Gamma}^\alpha_{\sigma}]_{jl}=2\pi
V_{j\alpha\sigma}V^*_{l\alpha\sigma}\rho_\alpha(\omega)$, and the
matrix elements are constants considering the two-dimensional nature
of the density of states $\rho_\alpha(\omega)$ in lead-$\alpha$. By
defining $\Gamma_0=2\pi|V|^2\rho_\alpha(\omega)$, we can write these
matrixes as
\begin{eqnarray}
&&[\bm{\Gamma}^L_\sigma]=\Gamma_0\left[\begin{array}{cc}
1 & e^{i\frac{\phi}{2}+i\sigma\varphi}\\
  e^{-i\frac{\phi}{2}-i\sigma\varphi} &1
\end{array}\right]\,,\nonumber\\
&&[\bm{\Gamma}^R_\sigma]=\Gamma_0\left[\begin{array}{cc}
1 & e^{-i\frac{\phi}{2}-i\sigma\varphi}\\
  e^{i\frac{\phi}{2}+i\sigma\varphi} &1
\end{array}\right]\,,
\end{eqnarray}
and
$[\bm{\Gamma}_{\sigma}]_{jl}=\frac{1}{2}([\bm{\Gamma}^L_\sigma]_{jl}+[\bm{\Gamma}^R_\sigma]_{jl})$
denotes the average QD-lead coupling. $\bm{G}^r_{\sigma}$ and
$\bm{G}^a_{\sigma}$, the retarded and advanced Green functions, obey
the relationship $[\bm{G}^r_{\sigma}]=[\bm{G}^a_{\sigma}]^\dag$. And
the matrix elements of $\bm{G}^r_\sigma$ are defined as
$[\bm{G}^r_\sigma]_{lj}(\omega)=\int^{+\infty}_{-\infty}[\bm{G}^r_\sigma(t)]_{lj}e^{i\omega
t}dt$ with $[\bm{G}^r_\sigma(t)]_{lj}=-i\theta(t)\langle
\{d_{l\sigma}(t),d^\dag_{j\sigma}\}\rangle$. The chemical potential
of lead-L $\mu_L$ is considered as the zero point of energy of the
system, denoted as $\mu_L=0$. $\mu_R$ and $\mu_D$, the chemical
potentials in other two leads, are fixed at
$\mu_{R}=\mu_D=\mu_L-\delta$. Thereby, the electron injects from
lead-L and departs from the QDs via the other two leads. Thus there
are two channels for the electronic tunneling in this structure,
i.e., the $L\rightarrow R$ channel and $L\rightarrow D$ channel.
When the electron transport is in the linear regime and at low
temperature, the current flow defined in Eq.(\ref{current}) reduces
to $J_{\beta\sigma}=\mathcal {G}_{\beta L,\sigma}[-\delta]$. So the
current flow in lead-$\beta$ is proportional to the linear
conductances
\begin{eqnarray}
\mathcal {G}_{\beta L,\sigma}=\frac{e^{2}}{h}T_{\beta
L,\sigma}(\omega)|_{\omega=0},\label{conductance}
\end{eqnarray}
with $\beta=R$ or $D$. It is obvious that the linear conductances
are associated with the Green functions, which can be solved by
means of the equation-of-motion method. By a straightforward
derivation, we obtain the matrix form of retarded Green function
\begin{eqnarray}
&&[\bm{G}_\sigma^r(\omega)]\nonumber\\
&&=\left[\begin{array}{cc} (z-\varepsilon_{1})S_{1\sigma}+i\Gamma_0 & i\Gamma_0\cos(\frac{\phi}{2}+\sigma\varphi)\\
  i\Gamma_0\cos(\frac{\phi}{2}+\sigma\varphi)& (z-\varepsilon_{2})S_{2\sigma}+i(\Gamma_0+\frac{\Gamma_D}{2})
\end{array}\right]^{-1}\ \label{green},
\end{eqnarray}
with $z=\omega+i0^+$ and $\Gamma_D=[\bm{\Gamma}^D_\sigma]_{22}$. And
$S_{j\sigma}=\frac{z-\varepsilon_{j}-U_j}{z-\varepsilon_{j}-U_j+U_j\langle
n_{j\bar{\sigma}}\rangle}$ accounts for the contribution of the
Coulomb interaction up to the second-order
approximation\cite{Gong,Liu}. The average electron occupation number
in QD is determined by the relations $\langle
n_{j\sigma}\rangle=-\frac{1}{\pi}\int d\omega \mathrm {Im}
[\bm{G}^{r}_\sigma]_{jj}$. So the Green function can be numerically
resolved by iteration technique.

\section{Numerical results and discussions \label{result2}}
With the formulation developed in the previous section, we can
perform the numerical calculation to investigate the linear
conductance spectra of the two-channel AB interferometer. Prior to
the calculation, we need to introduce a parameter $t_0$ as the units
of energy.
\par

First of all, we are concerned with the adjustment of a magnetic
field on the linear conductances. In Fig.\ref{flux}(a) the
conductances versus the magnetic phase $\phi$ is plotted in the
absence of Rashba interaction. We choose the coupling between QD-2
and lead-D $\Gamma_D=\Gamma_0=2t_0$, and consider the QD levels
$\varepsilon_1=\varepsilon_2=0$. In such a simple case, the
Hamiltonian is independent of the electron spin, hence
$\mathcal{G}_{\beta L,\sigma}=\mathcal{G}_{\beta
L,\bar{\sigma}}=\mathcal{G}_{\beta L}$. Besides, the Hubbard $U$
does not affect the electron transport since
$\varepsilon_jS_{j\sigma}=0$ according to the treatment of the
electron interaction given above. An interesting result shown in
Fig.\ref{flux}(a) is that the two conductances, $\mathcal {G}_{RL}$
and $\mathcal {G}_{DL}$, do not vary with $\phi$ in phase. In
particular, when $\phi=n\pi$, the magnetic phase being a multiple of
$\pi$, the peak of one conductance (${\mathcal G}_{RL}$ or
${\mathcal G}_{DL}$) just encounters the valley of another one.

\par
When Rashba interaction comes into play, the linear conductance
becomes spin dependent. In Fig.\ref{flux}(b) and (c) the linear
conductances versus the magnetic phase $\phi$ are plotted in the
presence of Rashba interaction with the corresponding phase
$\varphi=\frac{\pi}{4}$. Under the current experimental circumstance
such an SO coupling strength is available\cite{Serra}. From
Fig.\ref{flux}(b) and (c) we can see that the conductances $\mathcal
{G}_{\beta L,\sigma}$ and $\mathcal {G}_{\beta L,\bar{\sigma}}$ show
remarkable difference from each other around the points of
$\phi=(n+\frac{1}{2})\pi$, which implies a striking spin
polarization in the current flow in lead-$\beta$. More importantly,
$\mathcal {G}_{R L,\sigma}$ and $\mathcal {G}_{D L,\bar{\sigma}}$
almost oscillate in phase. At $\phi=(n+\frac{1}{2})\pi$, both reach
their respective maxima. This means that around the points
$\phi=(n+\frac{1}{2})\pi$ the incident electron from lead-L can
select a specific lead to leave the QDs according to its spin state.
Thus, we implement the spin polarization and spin separation
simultaneously in such a structure at a specific magnetic field. In
addition, by adjusting the magnetic field, the orientations of spin
polarization in lead-R and lead-D can be just interchanged.

\par
The underlying physics that gives rise to the spin dependent
electron transport property shown in Fig.\ref{flux} is quantum
interference. In order to obtain an intuitive picture about the
quantum interference, we analyze the electron transmission in the
language of Feynman path. To do so, we need firstly to rewrite the
electron transmission function in a form as
$T_{RL,\sigma}(\omega)=|\sum\limits_{j,
l=1}^2t_{RL,\sigma}(j,l)|^2$, where the electron transmission
probability amplitudes are defined as
$t_{RL,\sigma}(j,l)=\mathcal{V}_{Rj\sigma}G^r_{jl,\sigma}\mathcal{V}_{lL\sigma}$
with $\mathcal{V}_{j\alpha\sigma}=\mathcal{V}_{\alpha
j\sigma}^*=V_{j\alpha\sigma}\sqrt{2\pi\rho_\alpha(\omega)}$.
Following the expansion of the Green function
\begin{eqnarray}
G^r_{11,\sigma}=\frac{g^{-1}_{2\sigma}}{g^{-1}_{1\sigma}g^{-1}_{2\sigma}+\Gamma_{12,\sigma}\Gamma_{21,\sigma}}
=\sum\limits_{j=0}^\infty
g_1(-g_{1\sigma}g_{2\sigma}\Gamma_{12,\sigma}\Gamma_{21,\sigma})^j,\nonumber\\
\end{eqnarray}
we can then express the transmission probability amplitude
$t_{RL,\sigma}(1,1)$ as a summation of Feynman paths of all orders,
i.e.,
\begin{eqnarray}
&&t_{RL,\sigma}(1,1)\nonumber\\
&&=\sum\limits_{j=0}^\infty\mathcal{V}_{R1\sigma}
g_{1\sigma}(-g_{1\sigma}g_{2\sigma}\Gamma_{12,\sigma}\Gamma_{21,\sigma})^j\mathcal{V}_{1L\sigma}=\sum\limits_{j=0}^\infty
t^{(j)}_{RL,\sigma}(1,1).\nonumber\\
\end{eqnarray}
Here $g_{1\sigma}=(z-\varepsilon_1+i\Gamma_0)^{-1}$ and
$g_{2\sigma}=(z-\varepsilon_2+i\Gamma_0+\frac{i}{2}\Gamma_D)^{-1}$.
They are the Green functions of individual QD when another QD is
removed from the structure. It is obvious that higher-order Feynman
paths have more complicated forms. By the same token, we can expand
$t_{RL,\sigma}(1,2)$ as a summation of Feynman paths, which is given
by
\begin{eqnarray}
&&t_{RL,\sigma}(1,2)\nonumber\\
&&=\sum\limits_{j=1}^\infty
i\mathcal{V}_{R1\sigma}(-g_{1\sigma}g_{2\sigma}\Gamma_{12,\sigma})^{j}\Gamma_{21}^{j-1}\mathcal{V}_{2L\sigma}
=\sum\limits_{j=1}^\infty t^{(j)}_{RL,\sigma}(1,2).\nonumber\\
\end{eqnarray}
Besides, the other transmission coefficients $t_{RL,\sigma}(2,2)$
and $t_{RL,\sigma}(2,1)$ have the similar expansions as
$t_{RL,\sigma}(1,1)$ and $t_{RL,\sigma}(1,2)$. The lowest-order
Feynman paths in above equations are
$t^{(0)}_{RL,\sigma}(1,1)=\mathcal{V}_{R1\sigma}g_{1\sigma}\mathcal{V}_{1L\sigma}$
and
$t^{(0)}_{RL,\sigma}(2,2)=\mathcal{V}_{R2\sigma}g_{2\sigma}\mathcal{V}_{2L\sigma}$.
All Feynman paths contribute to the linear conductances coherently,
the lowest-order ones are the leading terms, though.
\par

As shown in Fig.\ref{zero-order}(a), the contribution of the
zero-order paths has the similar oscillation to the exact linear
conductance with the magnetic adjustment. Therefore, we can only
take into account the zero-order paths, i.e,
$t^{(0)}_{RL,\sigma}(1,1)$ and $t^{(0)}_{RL,\sigma}(2,2)$, to
analyze the quantum interference that cause the spin-dependent
electron transport between lead-R and lead-L, as shown in
Fig.\ref{flux}. The phase difference between the two zero order
paths is
$\Delta\theta_\sigma=[\phi+2\sigma\varphi+\theta_1-\theta_2]$, where
the phase $\theta_j$ arises from the Green function $g_{j\sigma}$,
i.e, $g_{j\sigma}=|g_{j\sigma}|e^{i\theta_j}$. According to the
above relation, we can realize that $\phi$, $\sigma\varphi$, and
$\theta_1-\theta_2$ are the phases associated with the magnetic
field, Rashba interaction, and the QD parameters, respectively. They
interplay to influence the quantum interference. And it should be
noticed that the Rashba interaction gives rise to the spin
dependence of the phase, hence to lead to the spin-dependent
conductance. For the cases shown in Fig.\ref{flux}, the Rashba
strength $\varphi=\frac{\pi}{4}$ and both $\theta_1$ and $\theta_2$
are equal to ${-\pi \over 2}$ due to QD levels being fixed at the
Fermi level. Then the change of $\phi$ influences the quantum
interference and brings out spin polarization. In the case of
$\phi=(n+\frac{1}{2})\pi$, the opposite-spin electrons will undergo
distinct quantum interference. For example, when
$\phi=\frac{1}{2}\pi$ the interference for spin-up electron between
the two zero-order Feynman paths is destructive with
$\Delta\theta_\uparrow=\pi$, but for the spin-down electron it is
constructive since $\Delta\theta_\downarrow=0$; However, in the case
of $\phi=\frac{-1}{2}\pi$ the situation just becomes opposite,
namely, the interference for spin-up electron is constructive
whereas it is destructive for the spin-down electron. In contrast,
in the case of $\phi=n\pi$ it can be readily seen that the quantum
interference and the conductance does not depend on the spin
freedom. Taking the case of $\phi=\pi$ as an example, one has
$\Delta\theta_{\uparrow}=\frac{3}{2}\pi$ and
$\Delta\theta_{\downarrow}=\frac{1}{2}\pi$. As a result, when the
opposite-spin electrons tunnel through the two paths, they undergo
the same quantum interference and no spin polarization occurs.

\par
With the similar method, we can analyze $T_{DL,\sigma}(\omega)$ by
writing out
$T_{DL,\sigma}(\omega)=|\sum\limits_{l=1}^2t_{DL,\sigma}(2,l)|^2$.
Here lead-R plays the same role as lead-D in the above case to act
on the quantum coherence. It is useful for us to give the explicit
forms of the lowest-order Feynman paths herein. They are
$t_{DL,\sigma}(2,2)$ and $t_{DL,\sigma}(2,1)$, with
Fig.\ref{zero-order}(b) shows the contribution of the above
lowest-order paths to the conductance, which presents the in-phase
oscillation to the exact conductance spectrum. So we can analyze the
quantum interference by taking only the lowest-order Feynman paths
into account. The phase difference between
$t^{(0)}_{DL,\sigma}(2,2)$ and $t^{(1a)}_{DL,\sigma}(2,1)$ is
$\nu_{0a,\sigma}=[\phi+2\sigma\varphi+\theta_1+{\pi \over 2}]$, the
phase difference between $t^{(0)}_{DL,\sigma}(2,2)$ and
$t^{(1b)}_{DL,\sigma}(2,1)$ is $\nu_{0b}=[\theta_1+{\pi \over 2}]$,
and the phase difference between $t^{(1a)}_{DL,\sigma}(2,1)$ and
$t^{(1b)}_{DL,\sigma}(2,1)$ is
$\nu_{ab,\sigma}=[\phi+2\sigma\varphi]$. With this phase relations,
we can find that, when $\phi=\frac{1}{2}\pi$,
$\nu_{0a,\uparrow}=2\pi$, $\nu_{0b}=\pi$, and
$\nu_{ab,\uparrow}=\pi$ which give rise to the constructive
interference for spin-up electron, but for the spin-down electron it
is destructive since $\nu_{0a,\downarrow}=\pi$ and
$\nu_{ab,\downarrow}=0$. Besides, with the help of these analysis,
it can also be found that the peak of ${\cal G}_{RL,\sigma}$
encounters the valley of ${\cal G}_{DL,\sigma}$ and vice versa.

\par
One of the characteristics of QD is its tunable level with respect
to the Fermi level, which can alter the phases of Feynman paths
taking part in the quantum interference. Therefore, we now turn to
discuss the variation of the spin-polarized current flow with the
shift of QD level. In the absence of a magnetic field but in the
presence of Rashba interaction with $\varphi=\frac{1}{4}\pi$, we
calculate the linear conductances as functions of
$\varepsilon_1=\varepsilon_2=\varepsilon_0$, which are shown in
Fig.\ref{gate}(a) and (b). From these spectra one can find the
notable and opposite spin polarizations in the $L\rightarrow R$ and
$L\rightarrow D$ channels except at the vicinity of
$\varepsilon_0=0$. Besides, in either channel the spin polarization
flips over when $\varepsilon_0$ passes through the Fermi level.
These results can be readily understood by analyzing the quantum
interference following the above argument. Namely, we will discuss
the quantum interference by taking only the lowest-order Feynman
paths into account. This is supported by the calculated results
shown in Fig.\ref{gate}(c) and (d), which are the contributions of
the lowest-order Feynman paths to the conductances. They show that
the spin polarizations in analogy with the exact ones shown in
Fig.\ref{gate}(a) and (b), respectively. As for the Feynman paths
from lead-R the phase difference
$\Delta\theta_\sigma=2\sigma\varphi+\theta_1-\theta_2$, in which $
\theta_1-\theta_2=\tan^{-1}\frac{\Gamma_0}{\varepsilon_0}
-\tan^{-1}\frac{2\Gamma_0+\Gamma_D}{2\varepsilon_0}$ is nonzero when
$\varepsilon_0$ departs from the Fermi level. As a result, when
electron tunnels through $t^{(0)}_{RL,\sigma}(1,1)$ and
$t^{(0)}_{RL,\sigma}(2,2)$, the phase difference is associated with
the electron spin. For example, at the point of $\varepsilon_0=2t_0
$ where the spin polarization is very striking, we can obtain
$\Delta\theta_\uparrow=\frac{7}{20}\pi$ and
$\Delta\theta_\downarrow=\frac{-13}{20}\pi$; Conversely, at the
point of $\varepsilon_0=-2t_0$,
$\Delta\theta_\uparrow=\frac{13}{20}\pi$ and
$\Delta\theta_\downarrow=\frac{-7}{20}\pi$. These distinct phase
differences just lead to opposite quantum interference for the
opposite-spin electrons in lead-R, hence the notable spin
polarization in the electron transport through this lead.
Furthermore, $\varepsilon_0$ passing through the Fermi level
inverses the spin polarization. As for the quantum interference in
the other channel, the situation is more complicated since there are
three low-order Feynman paths to be taken into account. We can find
$\nu_{ab,\uparrow/\downarrow}=(+/-)\frac{1}{2}\pi$ independent of
the adjustment of $\varepsilon_0$. Besides,
$\nu_{0b}=\frac{3}{4}\pi$, $\nu_{0a,\uparrow}=\frac{5}{4}\pi$, and
$\nu_{0a,\downarrow}=\frac{1}{4}\pi$, when $\varepsilon_0=2t_0$;
$\varepsilon_0=-2t_0$ corresponds to $\nu_{0b}=\frac{5}{4}\pi$,
$\nu_{0a,\uparrow}=\frac{7}{4}\pi$, and
$\nu_{0a,\downarrow}=\frac{5}{4}\pi$. The contribution of the three
low-order paths to the conductance is displayed in
Fig.\ref{gate}(d), which agrees with the calculations of these phase
differences. Based on these analysis, it is also clear that for the
same nonzero $\varepsilon_0$ the spin polarization in different
channels are opposite to each other. Therefore, with the tuning of
gate voltage, the spin-up polarized current in lead-R and spin-down
polarized current in lead-D can come into being simultaneously. It
is worth mentioning that the spin-dependent conductance spectra
shown Fig.\ref{gate} are obtained in the absence of a magnetic
field, which indicates that an external field is not indispensable
to cause the spin related quantum interference. Without a magnetic
field one can still fulfill the required result of the
spin-dependent electron transport by tuning the QD parameters.
\par
It should be pointed out that, the third lead(lead-D) is essential
to realize the spin polarization in the absence of a magnetic field.
For instance, when $\Gamma_D=0$, $g_{1\sigma}$ and $g_{2\sigma}$
correspond to the same phase($\theta_1-\theta_2\equiv0$), which
makes the total phase difference to be nothing to do with the
electron spin. Hence only in a multi-terminal structure the electron
transport with the spin-polarization effect can be fulfilled. And
for the two-terminal case and in the absence of magnetic field, spin
polarization is impossible.
\par
As is known, so far it is still a formidable challenge to fabricate
experimentally an QD structure consisting of identical QDs. Thereby
it is necessary for us to investigate the influence of the
fluctuations of structure parameters on the spin polarization and
separation during the electron transport process. We then calculate
the conductance spectra with the fluctuated structure parameters,
i.e, the QD levels and QD-lead couplings. The corresponding results
are shown in Fig.\ref{fluct}. From Fig.\ref{fluct}(a)-(b), we find
that the increase of the difference of QD levels brings out a little
weakening of the spin polarization when $\varepsilon_0$ is below the
Fermi level, whereas it strengthens the spin polarization for the
case of $\varepsilon_0$ above the Fermi level. However, the
fluctuation of QD levels can not result in the remarkable
destruction of the spin polarization and separation. On the other
hand, according to the results in Fig.\ref{fluct}(c)-(d) we can see
that the fluctuations of QD-lead couplings are not able to induce
the change of the electron transport in principle until the
fluctuations exceed $30\%$. Therefore, our conclusion is that the
spin polarization and separation in this structure does not require
an absolute uniformity of the structure parameters.
\par
Finally, we have to point out that the appropriate Rashba coupling
strength is crucial for the effect of spin polarization and
separation in such a three-terminal double-QD structure. To
illustrate this, we see the cases of zero magnetic field and
$\varphi=\frac{\pi}{2}$, thus
$\Gamma_{12,\sigma}=\Gamma_{21,\sigma}=0$. Accordingly, the
contributions of the higher-order paths become zero due to the
destructive quantum interference. In such a case
$T_{RL,\sigma}=|t^{(0)}_{RL,\sigma}(1,1)+t^{(0)}_{RL,\sigma}(2,2)|^2$,
and the phase difference resulting from the opposite-spin electrons
passing through these two paths is
$\Delta\theta_\sigma=[\pi+\theta_1-\theta_2]$, which is independent
of the electron spin. Alternatively,
$T_{DL,\sigma}=|t^{(0)}_{DL,\sigma}(2,2)|^2$, which is also
irrelevant to the spin index but shows a Breit-Wigner lineshape in
the linear conductance spectrum. Therefore, in this case the quantum
interference does not involve the electron spin, and no spin
polarization occurs.

\par
So far we have not discuss the effect of electron interaction on the
spin-dependent conductance spectra, though it is included in our
theoretical treatment. Now we incorporate the electron interaction
into the calculation of the conductance spectra. We wonder whether
it can destroy the property of spin polarization and separation. We
deal with the many-body terms by employing the second-order
approximation, since we are not interested in the electron
correlation here. And a uniform Coulomb repulsion $U=3t_0$ is
assumed for both QDs. Figure \ref{coulomb2} shows the calculated
conductance spectrum. It is found that in the linear conductance
spectra are separated into two groups due to the Coulomb repulsion,
and they are symmetric about the center of insulating band
$\varepsilon_0=-U/2$, which arises from the electron-hole
symmetry\cite{Gong}. Clearly, in such an approximation the
Rashba-related spin polarization and separation effect remain.

\par
With regard to the many-body effect, we should emphasize the
following point. Our calculation indicates that the electron
interaction does not destroy the spin polarization and separation
effect in this structure. However, we have taken the many-body terms
into account within the approximations only to second order and did
not consider the electron correlation, e.g, the Kondo effect. Thus
one can pay attention to the influence of Kondo resonance on the
spin polarization and separation by setting the QDs in the Kondo
regime. We would like to point out that in this system, the Kondo
effect can destroy the phenomena of spin polarization and
separation. It is known that in equilibrium, the influence of Kondo
effect on electron transport is to renormalize the QD levels to
coincide with the Fermi level of the system, which gives rise to the
Kondo resonance. However, according to the Feynman path theory above
such a property will bring out the invariability of the phase of
$g_{j\sigma}$ (i.e, $\theta_j\approx-\pi/2$), which will modify the
quantum interference in this system and restrain the spin
polarization and separation. Up to now, we can make a conclusion
that only in the case of relatively small Coulomb repulsion, the
properties of spin polarization and separation can be observed.
\par
We now turn to investigate the electron transport properties of this
structure in the case of finite bias voltage, to clarify the effect
of finite bias voltage on the spin polarization and separation. The
corresponding numerical results in Fig.\ref{bias} describe the
change of the current in the drain (lead-R and lead-D) with the
increase of bias voltage. In Fig.\ref{bias}(a), the noninteracting
case, we can see that with the adjustment of the bias voltage the
the properties of spin polarization and separation remain in this
electron transport process. Besides, when the many-body effect is
taken into account to the second order of the equation of motion,
the similar results to the noninteracting case are found, as shown
in Fig.\ref{bias}(b).

\section{summary}
In conclusion, we have theoretically investigated the electron
transport properties in a parallel double-QD structure with
three-terminals. By introducing the local Rashba spin-orbit
coupling, we find that an incident electron from one terminal can
select a specific terminal to depart from the QDs according to its
spin state. As a result, the functions of spin polarization and
separation can be simultaneously realized in this structure. And
spin polarizations in different channels can be inverted by tuning
the structure parameters. The underlying quantum interference that
gives rise to such results is analyzed in the language of Feynman
paths for the electron transmission. We find that the total phase
differences between any two low-order Feynman paths associated with
the Rashba interaction, the applied magnetic field, and the QD
scattering. In particular, the Rashba interaction provides a
spin-related phase, which is the origin of the spin-dependent
electron transport properties as we have reported. Besides, it
should be noted that a magnetic field and the QD parameters can
adjust the phase difference on the equal footing. Therefore, an
applied magnetic field is not indispensable to realize the spin
polarization in this structure. Instead of it, we can also obtain
these results via the adjustment of the QD parameters such as the QD
levels. In addition, by the detailed analysis on the quantum
interference, we find that for the appropriate Rashba interaction
strength(i.e., $\varphi\sim \pi/4$), it is possible to result in the
spin polarization and separation. And either of three leads is
absolutely necessary in the absence of magnetic field. However, for
a two-terminal structure it is impossible to obtain the spin
polarization without the magnetic field. When the approximation of
many-body effect is considered to second order, it is found that the
phenomena of spin polarization and separation remain. On the basis
of this feature, we propose that such a structure can be considered
as a device prototype to manipulate the spin freedom.

\clearpage

\bigskip
\begin{figure}
\caption{A schematic of the parallel double QD structure with three
terminals (labeled as L, R, and D). An electric field is applied to
QD-1 to induce the local Rashba interaction; $\Phi$ indicates a
magnetic flux penetrating the ring. \label{structure}}
\end{figure}

\begin{figure}
\caption{The linear conductances versus the magnetic phase $\phi$.
The structure parameters take the values as $\varepsilon_j=0$ and
$\Gamma_0=2t_0$. (a) In the absence of Rashba interaction. (b) and
(c) In the presence of Rashba interaction with the strength
$\varphi={1 \over 4}\pi$.\label{flux}}
\end{figure}

\begin{figure}
\caption{A comparison between the exact conductance and the
contributions of the lowest-order Feynman paths to the corresponding
conductances. \label{zero-order}}
\end{figure}

\begin{figure}
\caption{ (a) and (b) The linear conductances versus the QD level
$\varepsilon_0$ in the presence of Rashba interaction with
$\varphi={1 \over 4}\pi$. (c) and (d) The contribution of the
lowest-order Feynman paths. \label{gate}}
\end{figure}

\begin{figure}
\caption{ The conductance spectra in the presence of the fluctuated
QD parameters. (a) and (b) The conductances with the QD levels
different from each other. The results of
$\varepsilon_1=\varepsilon_0+\delta$ and
$\varepsilon_2=\varepsilon_0-\delta$ with $\delta=0$, $0.1t_0$, and
$0.5t_0$ are shown, respectively. (c) and (d) The conductances with
the fluctuation of the QD-lead couplings. The fluctuation is
measured by the quantity $\Delta_\Gamma=[\sum_j x_j^2/N-(\sum_j
x_j/N)^2]^{1/2}$ with $x_j=[{\bf{\Gamma}}^\alpha]_{jj}$, and the
cases of $\Delta_\Gamma=5\%$ and $=10\%$ are investigated.
\label{fluct}}
\end{figure}

\begin{figure}
\caption{The conductances versus the QD levels in the presence of
electron interaction with $U=3t_0$ in both QDs. \label{coulomb2}}
\end{figure}

\begin{figure}
\caption{The current versus the finite bias voltage. (a) corresponds
to the noninteracting case, and (b) is the result in the presence of
many-body effect. \label{bias}}
\end{figure}


\begin{thebibliography}{}

\bibitem{Das} I. \v{Z}uti\'{c}, J. Fabian, and S. Das Sarma, Rev. Mod. Phys. 6
(2004) 323.
\bibitem{science} S. A. Wolf, D. D. Awschalom, R. A. Buhrman, J. M. Daughton, S.
von Moln\'{a}r, M. L. Roukes, A. Y. Chtchelkanova, and D. M. Treger,
Science 294 (2001) 1488.
\bibitem{Das2} I. \v{Z}uti\'{c}, J. Fabian, and S. Das Sarma, Appl. Phys. Lett. 79 (2001)
1558.
\bibitem{Das3} I. \v{Z}uti\'{c}, J. Fabian, and S. Das Sarma, Appl. Phys. Lett.
82 (2003) 221.
\bibitem{Filip} G. Schmidt, D. Ferrand, L. W. Molenkamp, A. T. Filip, and B. J. V. Wees, Phys.
Rev. B 62 (2000) 4790.
\bibitem{Hbner} J. Hbner, W. W. Rhle, M. Klude, D. Hommel, R. D. R. Bhat, J. E. Sipe,
and H. M. van Driel, Phys. Rev. Lett. 90 (2003) 216601.

\bibitem{Rashba} A. Bychkov and E. I. Rashba,
J. Phys. C 17 (1984) 6039.
\bibitem{Rashba2} J. Nitta, T. Akazaki, H.
Takayanagi, and T. Enoki, Phys. Rev. Lett. 78 (1997) 1335.

\bibitem{Rashba3} G. Engels, J. Lange, Th. Sch\"{a}pers, and H. L\"{u}th, Phys. Rev. B
\textbf{55}, (1997) R1958; D. Grundler, Phys. Rev. Lett. 84 (2000)
6074.
\bibitem{Sun} Q. F. Sun, J. Wang, and H. Guo, Phys. Rev. B 71 (2005)
165310.
\bibitem{Datta} S. Datta and B. Das, Appl. Phys. Lett. 56 (1990) 665.

\bibitem{Kato} Y. K. Kato, R. C. Myers, A. C. Gossard, and D. D. Awschalom, Science
306 (2004) 1910.
\bibitem{Sinova} J. Wunderlich, B.
K\"{a}stner, J. Sinova, and T. Jungwirth, Phys. Rev. Lett. 94 (2005)
047204.

\bibitem{Xie} Q. F. Sun and X. C. Xie, Phys. Rev. B 73 (2006) 235301.
\bibitem{Serra} R. L¨®pez, D. Sanchez, and L. Serra, Phys. Rev. B 76 (2007) 035307.
\bibitem{Chi} F. Chi and S. S. Li, J. Appl. Phys. 100 (2006)
113703.

\bibitem{Meir1} Y. Meir and N. S. Wingreen, Phys. Rev. Lett. 68
(1992) 2512.
\bibitem{Meir2} A. P. Jauho, N. S. Wingreen, and Y. Meir, Phys. Rev. B
50 (1994) 5528.

\bibitem{Gong} W. Gong, Y. Zheng, Y. Liu, and T. L\"{u}, Phys. Rev. B 73 (2006)
245329.
\bibitem{Liu} Y. Liu, Y. Zheng, W. Gong, and T. L\"{u}, Phys. Lett. A 360 (2006) 154.

\end{thebibliography}
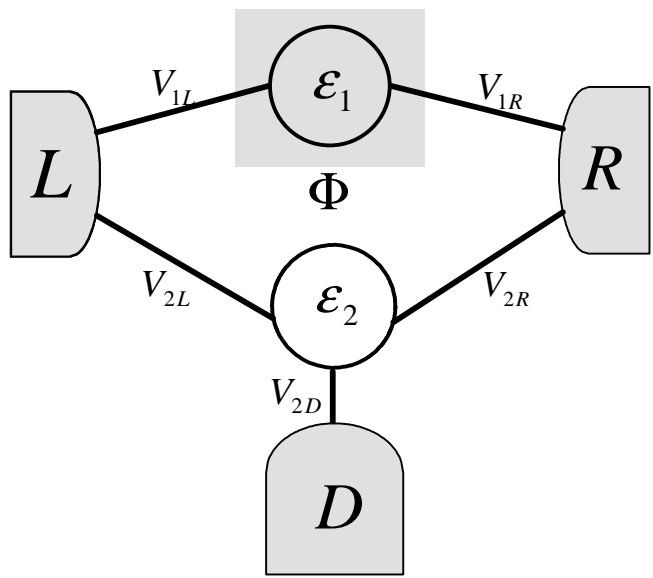
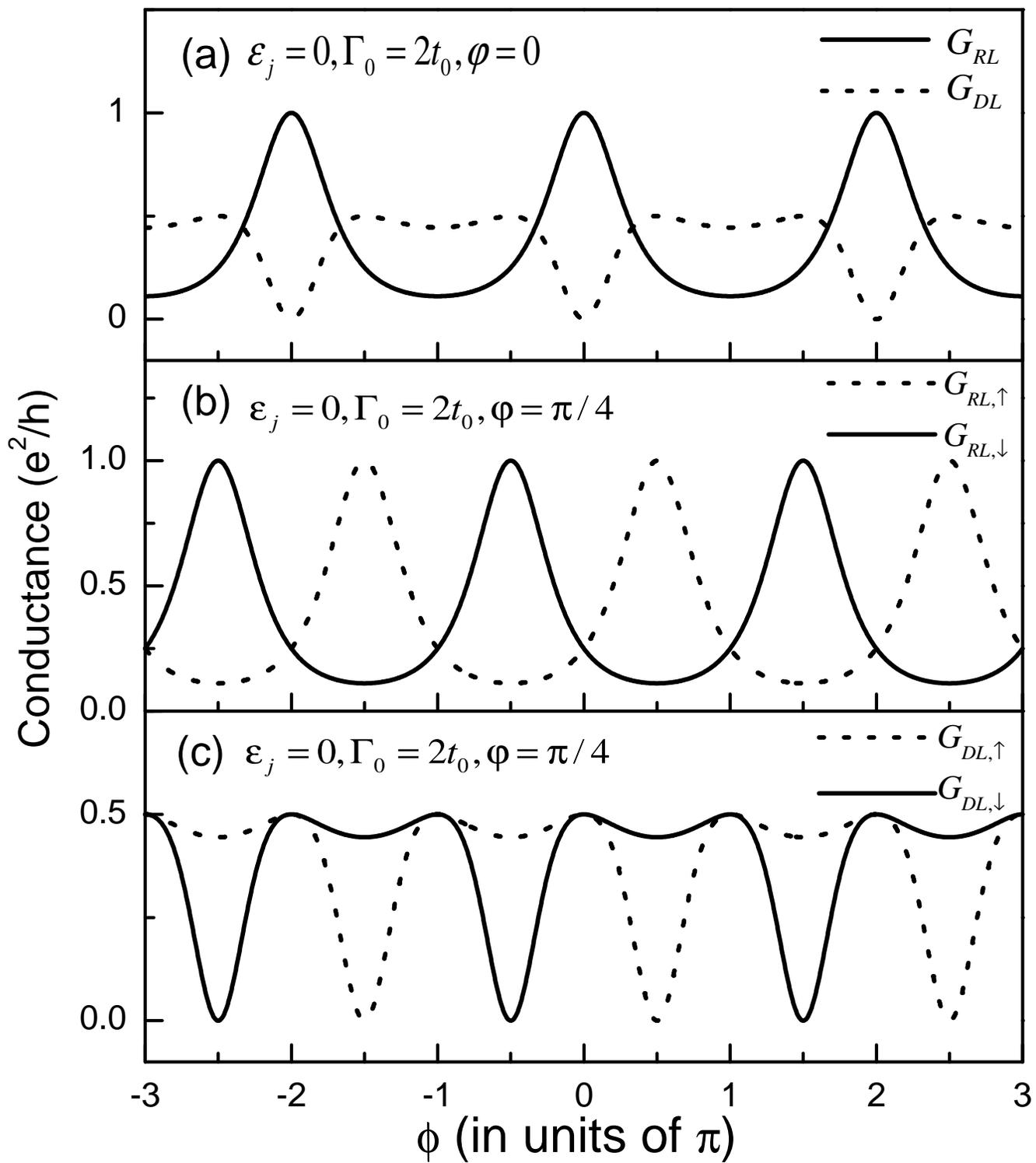
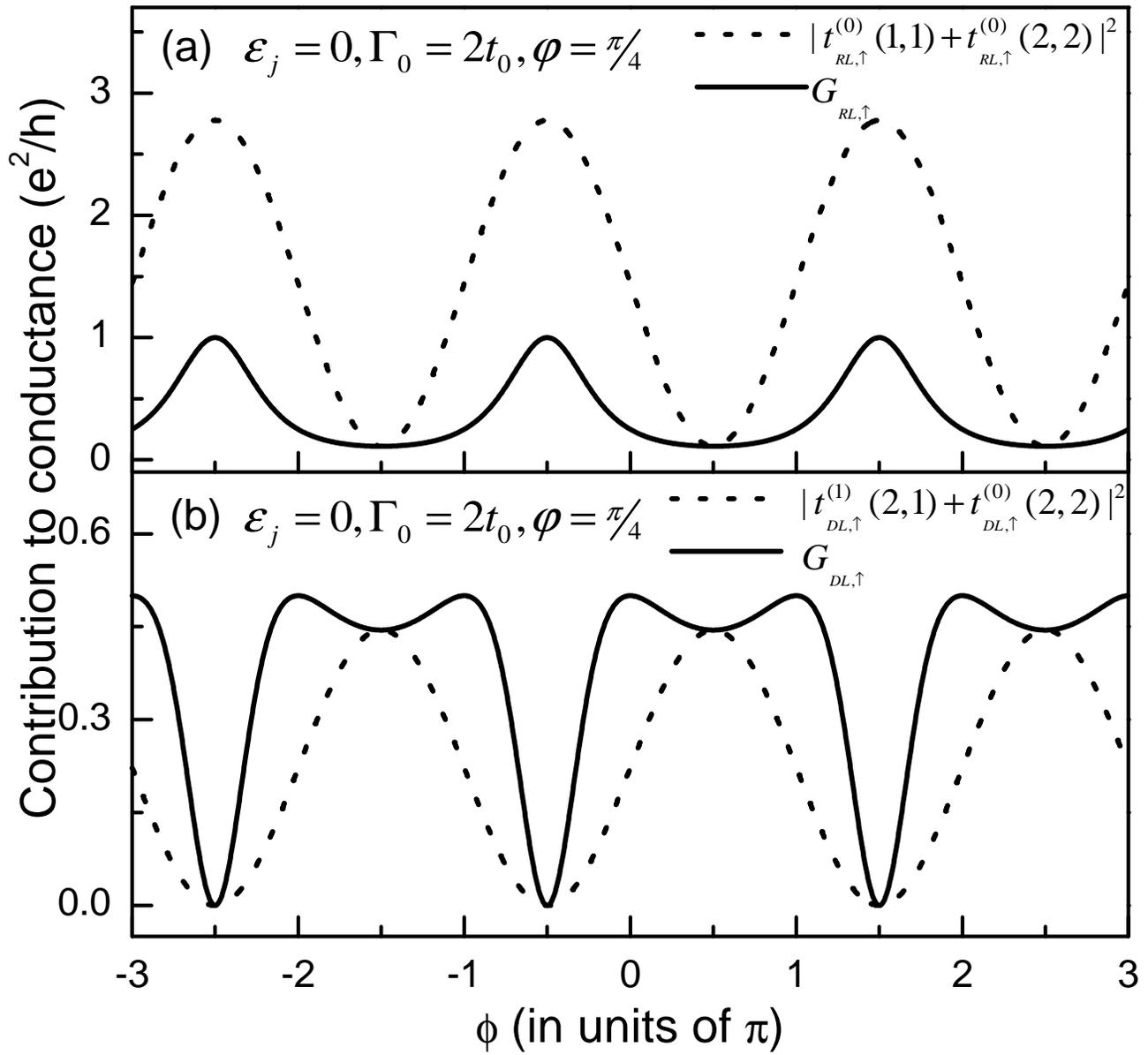
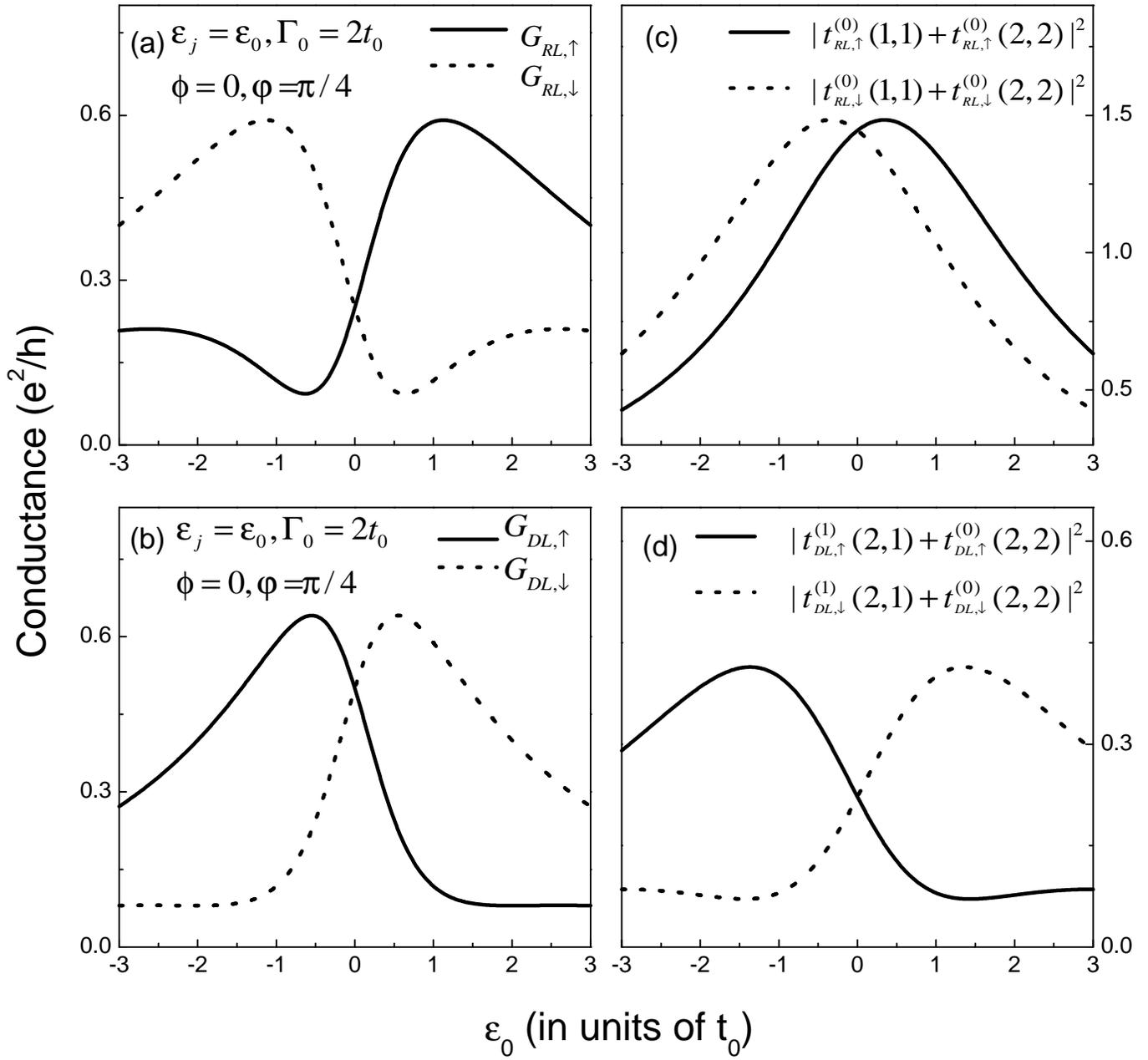
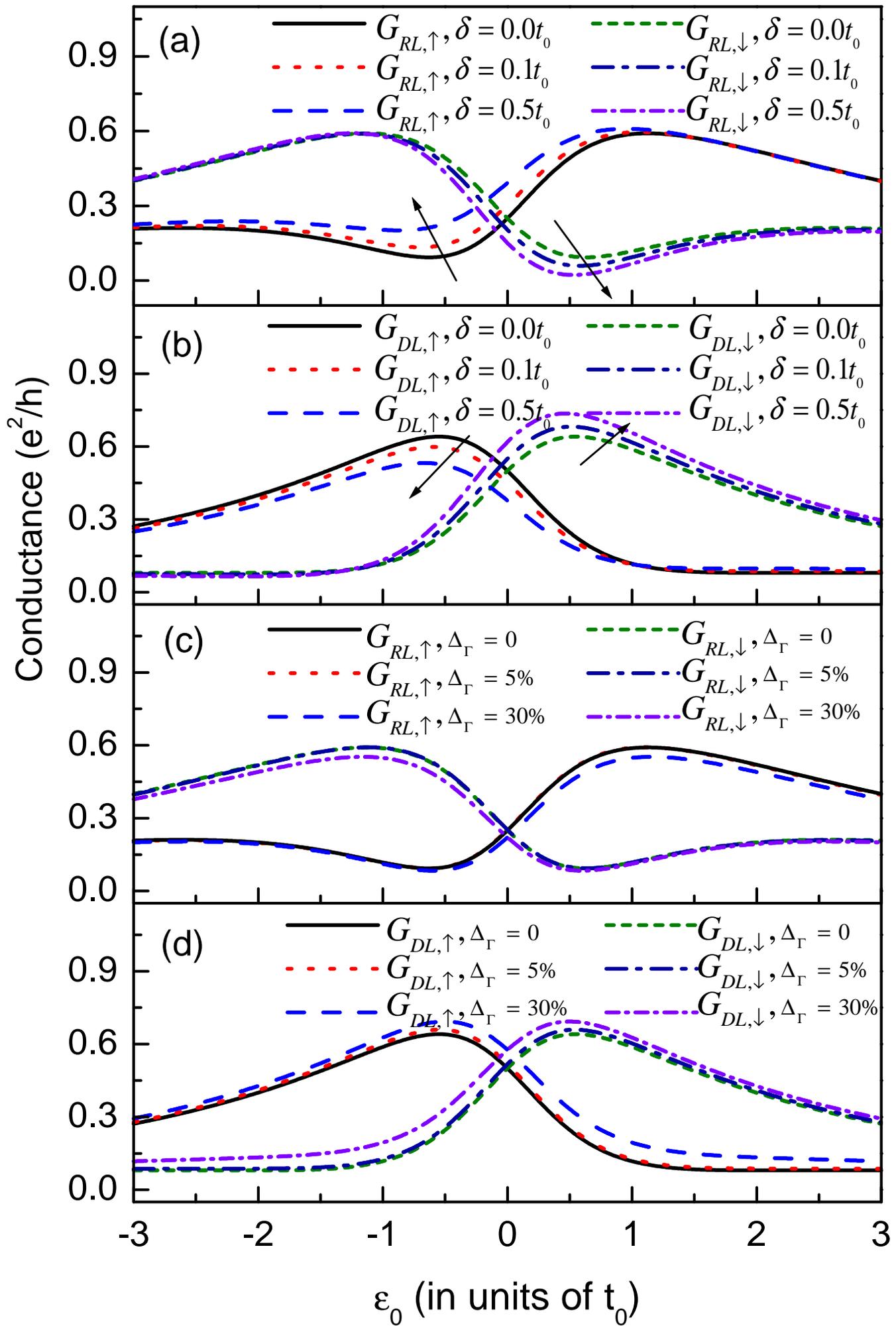
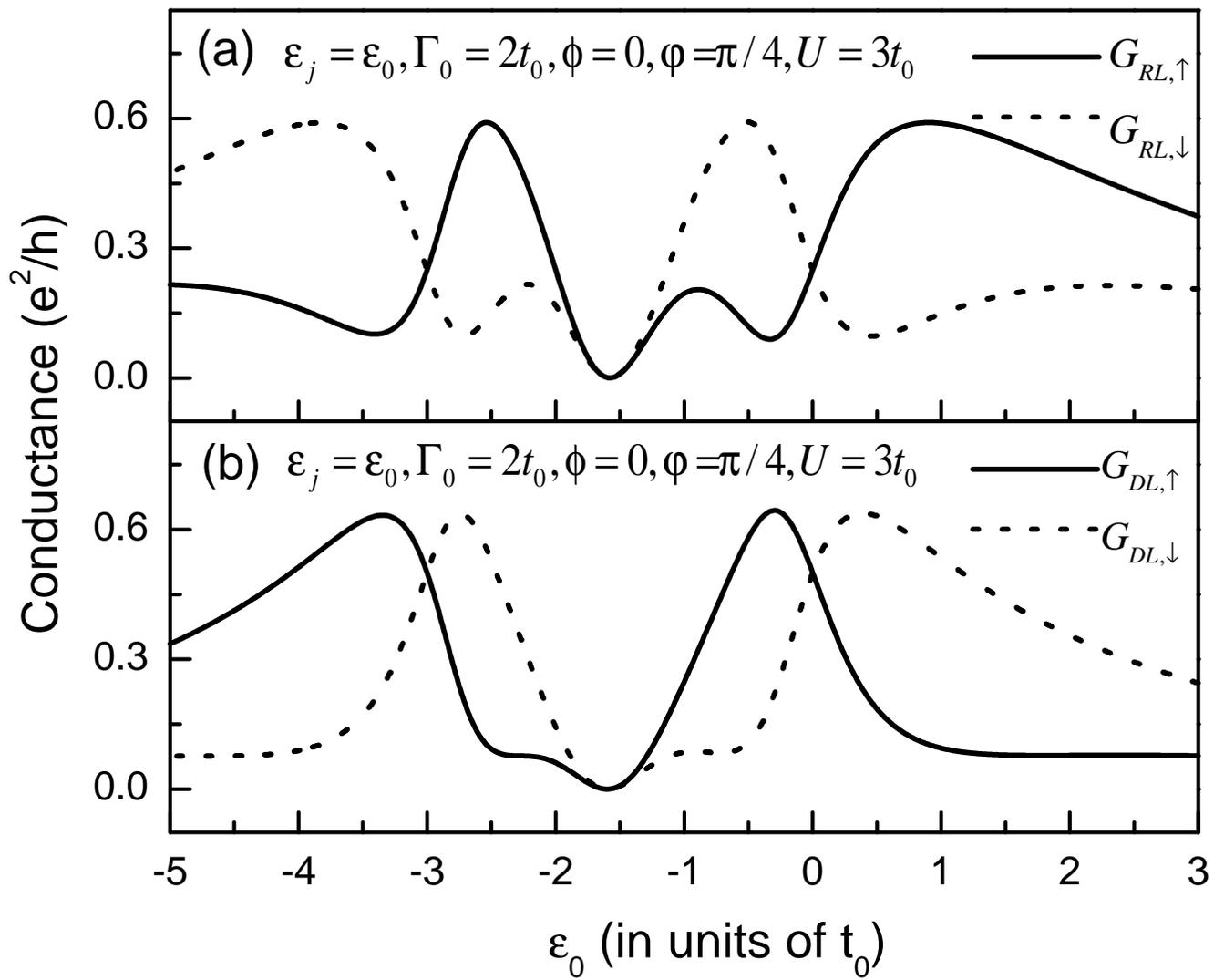
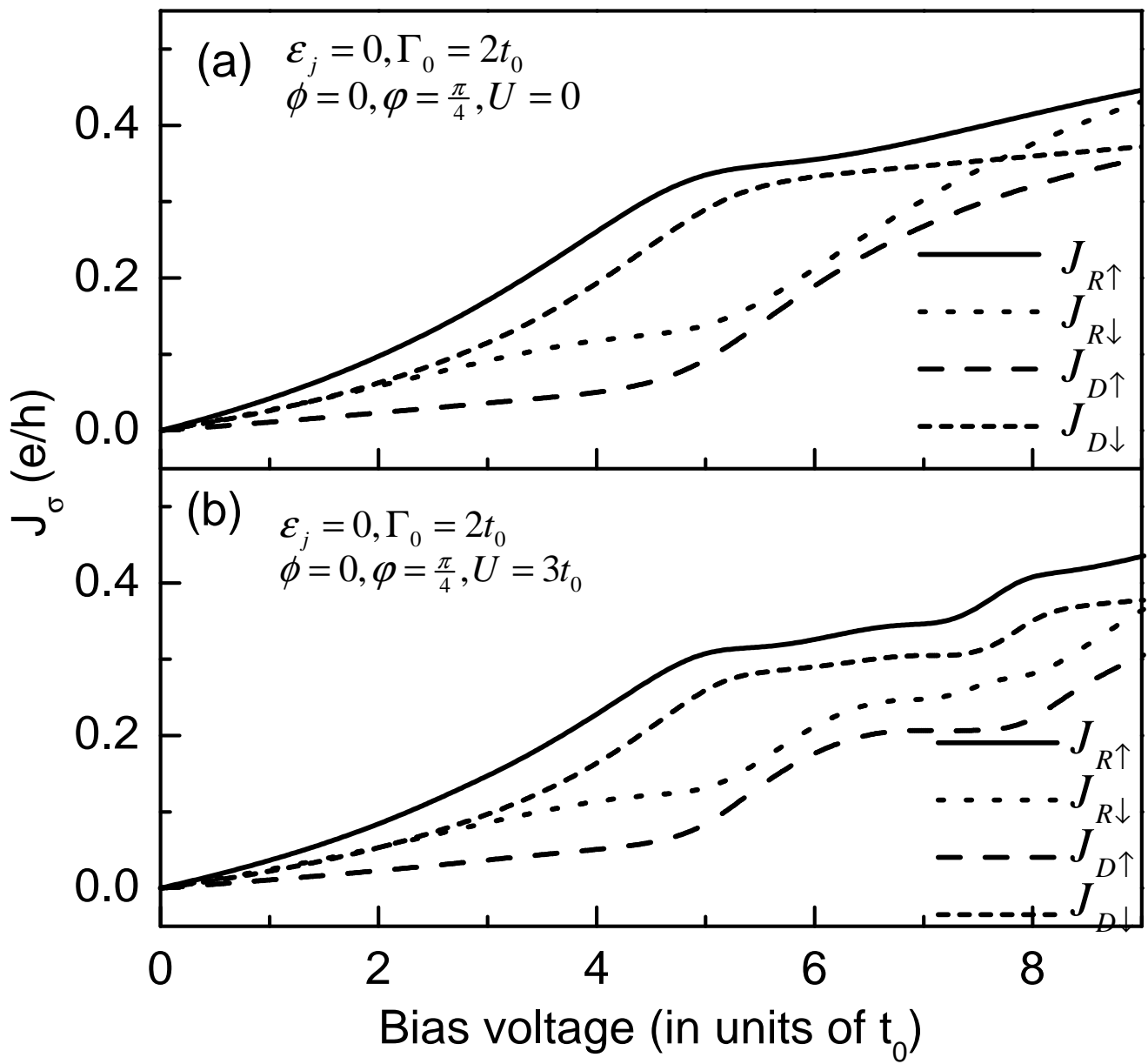
\end{document}